\documentclass[twocolumn,showpacs,preprintnumbers,amsmath,amssymb]{revtex4}
\def\rset{{\rm I\kern -0.2em R}} 
\begin{document}

\title{A Note on the Extension of the Polar Decomposition for 
the Multidimensional Burgers Equation} 
\author{
U.~Frisch$^{1,2}$ and M.~Mineev-Weinstein$^{3}$\\ \small{$^{1}$ CNRS,
Observatoire de la C\^ote d'Azur, B.P. 4229, 06304 Nice Cedex 4,
France.}\\ \small$^2$ CNLS, Theoretical Division, LANL, Los Alamos, NM 87545,
USA\\ \small$^3$ Applied Physics Division, Group X-7, MS-P365, LANL, Los 
Alamos, NM 87545, USA.}
\begin{abstract}
It is shown that the generalizations to more than one space
dimension of the pole decomposition for the Burgers equation
with finite viscosity $\nu$ and no force are  of the form ${\bf u} =
-2\nu \nabla \log P$, where the $P$'s are explicitly known algebraic (or
trigonometric) polynomials in the space variables with polynomial
(or exponential) dependence on time. Such solutions have polar singularities
on complex algebraic varieties.
\end{abstract}

\maketitle
\vspace{.2in}
\section{Introduction}
\label{s:introduction}

We are interested in the Burgers equation in $\rset^n$ \cite{Burgers}
\begin{equation}
\label{BurgersEq}
{\bf u}_t + {\bf u}\cdot \nabla {\bf u} = \nu \nabla^2 {\bf u}, \quad
{\bf u} = -\nabla \Phi,
\end{equation}
where ${\bf u}_t$ is the partial derivative of ${\bf u}$ with respect to time
$t$,
and $\nu$ is viscosity ($\nu >0$). This equation is the simplest evolutionary
dissipative equation, which is minimally (quadratically) nonlinear and enjoys
translational and Galilean invariance.  This simplicity and generality of the
equation (1) explains its applicability for seemingly different processes
occurring in a wide range of physical phenomena.  Although originally this
equation appeared as a model for Navier--Stokes turbulence \cite{Burgers}, 
it is mostly used today in cosmology \cite{applCosmology}, polymer physics 
\cite{applPolymers}, and nonlinear acoustics \cite{applAcoustics}.
Also this equation is very useful as a testing ground for numerical schemes in
hydrodynamics \cite{applHydrodynamics}. These 
features make the Burgers equation important and attractive for physicists.

From the mathematical point of view the Burgers equation is also remarkable, 
for it is completely integrable \cite{AblowitzNewell}, i.e.\ reducible to a
{\it linear} problem after equivalent transformation.  
This follows directly from the Cole-Hopf
transformation \cite{ColeHopf}:

\begin{equation}
\label{ColeHopf}
{\bf u} = -2\nu {\bf \nabla} \log \theta,
\end{equation}
which maps  (\ref{BurgersEq}) into the linear heat equation for the scalar field
$\theta$, namely

\begin{equation}
\label{HeatEq}
\theta_t = \nu \nabla^2 \theta,
\end{equation}
from which the  solution to the Burgers equation (\ref{BurgersEq}) can be 
obtained explicitly by quadrature.

With a few exceptions, completely integrable PDEs are two dimensional
(one dimension for time and one dimension for space) \cite{AblowitzNewell}.
There is a probable underlying mathematical reason, which hinders 
integrability in 
more than one spatial dimension: loosely speaking, it is related to the fact 
that polynomials with respect to more than one variable generally are not 
factorizable into non-trivial factors, while in the case of one variable they 
always are, by virtue of the main theorem of algebra.  However, the Burgers 
equation (\ref{BurgersEq}) is integrable
in arbitrary number of dimensions. 
This is obvious, because the Cole-Hopf transformation (\ref{ColeHopf}), 
which maps (\ref{BurgersEq})
to (\ref{HeatEq}), is valid in $\rset^n$ for an arbitrary natural $n$.  
Our goal here is 
to describe two classes of finite-dimensional exact solutions of the 
multidimensional Burgers equation (\ref{BurgersEq}), which are extensions of the 
``pole decomposition'' of the (1+1)-dimensional Burgers equation 
\cite{ChoodnChoodn}, \cite{FrischPoleDecomp}.

The pole decomposition is a property of PDEs (or integro-PDEs)
to have finite dimensional solutions whose degrees of freedom are movable 
singularities (poles) in the complex plane.  Most, if not all, completely 
integrable 
models enjoy this property \cite{AblowitzNewell}. 
The most notable examples of pole decomposition in integrable systems can be 
found
in Refs. \cite{ChoodnChoodn}, \cite{Kruskal74}-\cite{Moser}.
As said above, the Burgers equation (\ref{BurgersEq}) in  (1+1)-dimension also enjoys the pole 
decomposition \cite{ChoodnChoodn}, \cite{FrischPoleDecomp}. (See also 
\cite{Kimura}, \cite{Senouf} for recent results in this direction.)
Namely, it admits ``polar'' solutions in the form
\begin{equation}
\label{BurgPoleDecomp}
{\bf u}(t,x) = -2\nu \sum_{k=1}^N \frac{1}{x - z_k(t)},
\end{equation}
where  the poles constitute an $N$-dimensional dynamical system:
\begin{equation}
\label{PoleDynamics}
\frac{d z_l}{d t} = -2\nu \sum_{k \neq l}^N \frac{1}{z_l - z_k},
\end{equation}

The pole decomposition for (\ref{BurgersEq}) corresponds to solutions 
of (\ref{HeatEq}) which are 
polynomial in the space variable. Note that  existence of a pole
decomposition for a nonlinear system does not imply its 
integrability. Indeed, there are known instances of nonlinear {\it
non-integrable} models, 
which also possess  a pole decomposition.  They include the 2-D Euler
equation for ideal hydrodynamics \cite{Sommerfeld}, some models in plasma 
turbulence \cite{LeeChen}, some versions of the Sivashinsky equation for 
a flame propagation \cite{ThualFrischHenon}, \cite{ProcacciaOlami}, and 
related combustion systems \cite{GuyJoulin}.
Both integrable and nonintegrable systems possessing a pole decomposition are
of interest to  physicists and mathematicians, for the pole dynamics
reveals important physical trends and hidden mathematical
structure underlying the model.

It should also be mentioned that the Burgers equation  is dissipative,
unlike almost all integrable systems, which are Hamiltonian \cite{AblowitzNewell}.
The phase volume of closed dissipative systems shrinks with time, so that only
a few degrees of freedom are really relevant in the long-time asymptotics,
most of the initially existing degrees of freedom being eventually 
suppressed.
Because a pole decomposition is an exact finite-dimensional reduction of the 
system with an infinite number of degrees of freedom, this explains why such a
decomposition is especially instructive for dissipative models, like the
Burgers equation (\ref{BurgersEq}) \cite{FrischPoleDecomp}, flame propagation 
\cite{ThualFrischHenon}-\cite{GuyJoulin}, and viscous fingering 
(the Saffman-Taylor problem) \cite{LGE}.

Since multidimensional integrability is a far more difficult subject than
the $(1+1)$-dimensional case, it is tempting to extend to higher dimensions
physical ideas adopted from interacting poles, by replacing poles by strings
or by more general complex varieties. A priori however it seems impossible to
extend (\ref{PoleDynamics}) to non-pointlike objects, while keeping a finite 
number of degrees
of freedom. However we shall see that in higher dimensions there are still
polynomial (algebraic and trigonometric)  solutions to (\ref{HeatEq}) which can 
be obtained explicitly.
Observe that polynomials are factorizable in one dimension, whereas this is
generally not the case in higher dimensions: the zeros of polynomials are
then located on algebraic  complex varieties which are generally 
{\it irreducible\/} \cite{AlgGeom}.

Here we will show how to construct polynomial-based solutions to the multidimensional Burgers equation 
(\ref{BurgersEq}) with singularities on such irreducible varieties

\vspace{.2in}

\section{Solutions generated by polynomials}
\label{s:polsol}

We are looking for polynomial solutions to the heat  equation
\begin{equation}
\label{AlgPolyn}
P(t, {\bf x}) 
= \sum_{{\bf K = 0}}^{\bf M} a_{{\bf K}}(t) \prod_{l=1}
^n x_l^{k_l},
\end{equation}
where ${\bf x} = (x_1, x_2, ..., x_n)$, ${\bf K} = (k_1, k_2, ...,
k_n)$, ${\bf 0} = (0, 0, ..., 0)$, ${\bf M} = (m_1, m_2, ..., m_n)$, and 
$|{\bf M}| = \sum_{k=1}^n m_k$ is the degree of the polynomial.  It is 
technically convenient to define new coefficients
\begin{equation}
\label{badef}
b_{{\bf K}} = a_{{\bf K}}\,\prod_{l=1}^n k_l!\quad.
\end{equation}
The initial ($t=0$) values of these coefficients
are denoted by the superscript zero. 

It is easily checked by substitution into the heat equation that the
time dependence of the $b_{{\bf K}}$'s is
\begin{equation}
\label{AlgPolynDynamics}
b_{{\bf K}}(t) = \sum_{{\bf P = 0}}^{{\bf M'}} b_{{\bf K + 2P}}^0 
\frac{(\nu t)^{p_1 + p_2 + ... + p_n}}{(p_1 +p_2 + ... + p_n)!},
\end{equation}
where ${\bf M'} = (m_1', m_2', ..., m_n')$, and $m_l'$ is either $m_l$
or $m_l - 1$ depending on whether $m_l - k_l$ in
(\ref{AlgPolynDynamics}) is even or odd.

Using  (\ref{ColeHopf}) we find that our polynomial solutions
generate the following rational solutions to the Burgers equation
\begin{equation}
\label{BurgAlgSolutions}
u_p(t, {\bf x}) = - 2\nu \frac{\sum_{{\bf K} = 0}^{\bf M}
a_{{\bf K}}(t) k_p
\prod_{l=1}^n x_l^{k_l}}{x_p \sum_{{\bf K} = 0}^{\bf M} 
a_{{\bf K}}(t) \prod_{l=1}^n x_l^{k_l}},
\end{equation}
where $u_p$ is the $p^{th}$ component of the vector field ${\bf u}$.
Another closely related class of solutions involves trigonometric
polynomials
\begin{equation}
\label{TrigPolyn}
P(t, {\bf x}) = {\cal R}e \sum_{{\bf K} = 0}^{\bf M} c_{{\bf K}}(t) 
\prod_{l=1}^n e^{i k_l x_l}
\end{equation}
with the same ${\bf K}$ and ${\bf M}$ as in (\ref{AlgPolynDynamics}). The 
time-dependence of $c_{{\bf K}}$ is now given by  
\begin{equation}
\label{TrigPolynDynamics}
c_{{\bf K}} (t)= c_{{\bf K}}^0 \exp\left(- \nu 
\left(\sum_{l=1}^n k_l^2\right)t\right).
\end{equation}
By (\ref{ColeHopf}) this generates the following solutions 
to the Burgers equation 
\begin{equation}
\label{BurgTrigSolutions}
u_p(t, {\bf x}) = 2 \nu \frac{{\cal I}m \sum_{{\bf K} = 0}^{\bf M} 
c_{{\bf K}}(t) \, k_p
\prod_{l=1}^n e^{i k_l x_l}}{{\cal R}e \sum_{{\bf K} = 0}^{\bf M} 
c_{{\bf K}}(t)
\prod_{l=1}^n e^{i k_l x_l}}.
\end{equation}

\vspace{.2in}

Thus we have shown that the Burgers equation (\ref{BurgersEq}) in $\rset^n$
possesses exact solutions with a finite number of time dependent
parameters generated by the algebraic and trigonometric polynomial
solutions of the heat equation in $\rset^n$.  

We observe that such solutions, contrary to the one-dimensional case,
cannot in general be decomposed into a sum of separate simpler
solutions. Indeed, this would correspond to having {\it at all time\/}
polynomial solutions of the heat equation which are factorized. Even
if the initial polynomial is
factorized, the time evolution will in general destroy the
factorization.  Recently,  special solutions possessing the
all-time factorization property were found  by D.~Leshchiner and one
of the authors (M.~M-W.). We do not yet know how broad is the class
of such solutions.

A final remark concerns integrability and explicit characterization of
singularities. Knowing explicitly the coefficients of the polynomial
solution of the heat equation does not imply that we can explicitly
describe the algebraic variety on which the polynomial vanishes. Even
in one dimension, if we have a pole decomposition with more than four
poles, we conjecture that Galois theory implies the following: given
the initial position, in general it is not possible to find the
positions for all times by radicals.

We gratefully acknowledge helpful discussions with F.~Calogero,
M.~Kruskal, and D.~Leshchiner.  This work was supported by the 
the LDRD project 2002006ER 
``Unstable Fluid-Fluid Interfaces'' at LANL, by the
European Union under contract HPRN-CT-2000-00162, and by the Indo-French Centre
for the Promotion of Advanced Research (IFCPAR~2404-2).

\end{document}